\documentclass[useAMS,usenatbib,usegraphicx]{mn2e}     
\usepackage{amsmath}
\usepackage{amssymb}
\usepackage{mathrsfs}
\usepackage{rotating}
\usepackage{longtable}
\usepackage{txfonts}
\usepackage{bm}

\newcommand{\beq}{\begin{equation}}
\newcommand{\eeq}{\end{equation}}

\newcommand{\ov}{\overline{{\rm v}}}
\newcommand{\kms}{\mathrm{km \: s^{-1}}}
\newcommand{\Vmax}{\mathrm{V}_\mathrm{max}}

\begin{document}

\title[Bayesian Object Detection]{Evidence for Substructure in Ursa Minor Dwarf Spheroidal Galaxy using a Bayesian Object Detection Method}

\author[A. B. Pace et al.]
{Andrew B. Pace$^1$, 
Gregory D. Martinez$^{1,2}$, 
Manoj Kaplinghat$^1$,
Ricardo R. Mu\~noz$^{3,4}$\\
$^1$Center for Cosmology, Department of Physics and Astronomy, University of California, Irvine, CA 92697\\
$^2$The Oskar Klein Center, Department of Physics, Stockholm University, Albanova, SE-10691 Stockholm, Sweden\\
$^3$Departamento de Astronom\'ia, Universidad de Chile, Casilla 36-D, Santiago, Chile\\
$^4$Department of Astronomy, Yale University, New Haven, CT 06520\\
}

\date{\today}

\maketitle

\begin{abstract}
{
We present a method for identifying localized secondary populations in stellar velocity data using Bayesian statistical techniques. We apply this method to the dwarf spheroidal galaxy Ursa Minor and find two  secondary objects in this satellite of the Milky Way. One object is kinematically cold with a velocity dispersion of $4.25 \pm 0.75\ \kms$ and centered at $(9.1\arcmin \pm 1.5, 7.2\arcmin \pm 1.2)$ in relative RA and DEC with respect to the center of Ursa Minor. The second object has a large velocity offset of $-12.8^{+1.75}_{-1.5}\  \kms$ compared to Ursa Minor and centered at $(-14.0\arcmin^{+2.4}_{-5.8}, -2.5\arcmin^{+0.4}_{-1.0})$. The kinematically cold object has been found before using a smaller data set but the prediction that this cold object has a velocity dispersion larger than $2.0\ \kms$ at 95\% C.L. differs from previous work. We use two and three component models along with the information criteria and Bayesian evidence model selection methods to argue that Ursa Minor has one or two localized 
secondary populations. The significant probability for a large velocity dispersion in each secondary object raises the intriguing possibility that each has its own dark matter halo, that is, it is a satellite of a satellite of the Milky Way.\\
}
{{\sc keywords: Dark Matter: Substructure, Dwarf Galaxies: Ursa Minor, Bayesian Statistics}}
\end{abstract}

\section{Introduction}

The Milky Way dwarf spheroidal galaxies (dSphs) are the faintest but most numerous of the Galactic satellites.   
About 22 dSphs have been discovered with nine known before the Sloan Digital Sky Survey (SDSS).
The latter satellites
are often collectively referred to as the classical dSphs.
Thus, thanks to the advent of the SDSS, the number of known Milky Way dSphs has more than doubled 
\citep{Willman05, Belokurov06, Zucker06a, Zucker06b, Belokurov07, Sakamoto06, Irwin07, Walsh07}.
The classical systems are in general brighter and more extended than their post-SDSS
counterparts, usually referred to as the  ultra-faint dwarfs.
The dSph population of the Milky Way have a wide range of luminosities, 
$10^{3-7} L_{\odot}$, and sizes (half-light radii) from 40 to 1000 pc \citep{Mateo98, Simon07,Martin08}, 
but span a narrow range of dynamical mass: $M(r < 300 \rm pc) \approx 10^7$ M$_{\odot}$
for most of the dwarfs \citep{Strigari08}. In the context of hierarchical structure formation scenario, 
these dSphs would reside in the dark matter subhalos of the Milky Way host halo and so the dynamical mass provides an estimate of the amount of dark matter in subhalos. 
The dynamical mass-to-light ratios span a large range of 8-4000 (in solar units); 
some of these systems are the most dark matter dominated systems known \citep{Walker09c,Wolf10,Simon11,Martinez11}. 

Simulations also predict that subhalos should have their own subhalos (``sub-subhalos", e.g., \citealt{Shaw07, Kuhlen08, Springel08, Diemand08}). While their presence in cold dark matter simulations has been verified, the mass function of these sub-subhalos hasn't been well-quantified. The subhalo mass function is seen to follow a universal profile when scaled to the virial mass of the host halo. If the sub-subhalos follow the same pattern, then we expect to see a sub-subhalo with $\Vmax\simeq 0.3 \Vmax(\mathrm{subhalo})$ \citep{Springel08}.  We are motivated by this fact to search for stellar content that could be associated with these sub-subhalos.

Several dSphs show signs of stellar substructure or multiple distinct chemo-kinematic populations (Fornax, Sculptor, Sextans, Ursa Minor, Canes Venatici I).
For instance, in Fornax, there are stellar over-densities along the minor axis, possibly remnants of past mergers \citep{Coleman04, Coleman05} and five globular clusters \citep{Mackey03}.
In addition, Fornax's metal-rich and metal-poor stellar components seem to have different velocity dispersions \citep{Battaglia06}.
Similarly,  Sextans and Sculptor  each contain two kinematically distinct secondary populations with different metallicities \citep{Bellazzini01, Battaglia08}. Sculptor's populations have different velocity dispersion profiles, in addition to their distinct metalicities \citep{Battaglia08}, whereas Sextans has localized kinematically distinct population either near its center \citep{Kleyna04, Battaglia11} or near its core radius \citep{Walker06}.
There are claims of two populations with distinct velocity and metallicity distributions in the brightest ultra-faint dwarf, Canes Venatici I (CVI) \citep{Ibata06}, but this is not seen in two other data sets  \citep{Simon07, Ural10}.
The Bo\"{o}tes I ultra-faint could also have two kinematically distinct populations with different scale lengths \citep{Koposov11}, although this wasn't apparent in earlier data sets \citep{Munoz06-bootes, Martin07}. 
The largest of these Bo\"{o}tes I data sets contains 37 member stars and this has to be weighed against the results of \citet{Ural10} who suggest that at least 100 stars are required to differentiate two populations. 

\begin{table}
\label{tab:umi}
\caption{Observed and derived properties of Ursa Minor.}
\begin{tabular}{lr}
Parameter & Value \\
\hline \hline
Distance$^{\, 1}$ & $77 \pm 4 $ kpc \\
Luminosity$^{\, 1}$  & $3.9^{+1.7}_{-1.3} \times 10^5 {\rm L}_{\odot,{\rm V}}$\\
Core radius$^{\, 1}$ & $17.9\arcmin \pm 2.1$\\
Tidal radius$^{\, 1}$ & $77.9\arcmin \pm 8.9$\\
Half-light radius$^{\, 1}$  & $0.445 \pm 0.044$ kpc \\
Deprojected half-light radius$^{\, 1}$ ($r_{1/2}$) & $0.588 \pm 0.058$ kpc\\
Average velocity dispersion$^{\, 2}$ & $11.61 \pm 0.63 \: \kms$ \\
Mean velocity$^{\, 2}$ & -247 km/s\\
Dynamical mass within $r_{1/2}$$^{\, 1}$ & $5.56^{+0.79}_{-0.72} \times 10^7 M_{\odot}$ \\ 
Mass-to-light ratio within $r_{1/2}$$^{\, 1}$ & 290$^{+140}_{-90}  M_{\odot}/ {\rm L}_\odot$\\
Ellipticity$^{\, 3}$ & $0.56 \pm 0.05 $ \\
Center (J2000.0)$^{\, 4}$ & $(15^h09^m10^{s}.2, 67^{\circ}12'52'')$\\
Position angle$^{\, 5}$ & $49.4^{\circ}$\\
\hline
\end{tabular}\\
Note: References are as follows 1. \cite{Wolf10} and references therein 2. This paper  3. \cite{Mateo98} 4. \cite{Kleyna03} 5. \cite{Kleyna98}\\
\end{table}

Among the classical dSphs, only Draco has a lower V-band luminosity but Ursa Minor is twice as extended as Draco (in terms of its half-light radius) \citep{Irwin95, Palma03}. 
Its observed and derived properties are summarized in Table~\ref{tab:umi}. Ursa Minor is also likely the most massive satellite in terms of its dark matter halo, apart from the Magellanic clouds and the disrupting Sagittarius dSph. These properties make Ursa Minor an ideal target to search for substructure.
The $\Vmax$ at infall for the subhalo hosting Ursa Minor should be greater than 25 km/s but probably no larger than about 50 km/s \citep{Boylan_Kolchin2012} and thus we can 
expect Ursa Minor to have a sub-subhalo with $\Vmax$ in the range of $8-16 \kms$. Despite its low mass, such a small sub-subhalo could have held on to its gas because it was protected by the deeper potential well of Ursa Minor.

Several photometric studies with different magnitude limits and overall extent observed, have reported additional localized stellar components of the stellar distribution that deviates from a smooth density profile \citep{Olszewski85, Kleyna98, Palma03}, particularly near the center \citep{Demers95, Eskridge01}.
To the northwest of the center, a secondary peak in the spatial distribution is seen in contours and isopleths \citep{Irwin95, Kleyna98, Bellazzini02, Palma03}.
However, different studies have concluded that this secondary peak is inconclusive or of low significance \citep{Irwin95, Kleyna98, Bellazzini02, Palma03}.
Smaller scale stellar substructure is, however, seen with higher significance \citep{Eskridge01, Bellazzini02}.
Combining proper motion information with shallow photometric data in the central 20 arcmin of Ursa Minor, 
\citet{Eskridge01} claim that the distribution of stars in Ursa Minor shows high significance 
for substructure in clumps of $\sim 3\arcmin0$ in size.
\citet{Bellazzini02} used the presence of a secondary peak in the distribution of the distance to the 200th neighboring star to argue that the surface density profile of Ursa Minor is not smooth. 
In addition, the stellar density is not symmetric along the major axis with the density falling more rapidly on the Western side \citep{Eskridge01, Palma03}
Statistically significant S-shaped morphology is also seen in contours of the red giant branch stars \citep{Palma03}.

Spectroscopic studies of Ursa Minor \citep{Hargreaves1994, Armandroff1995, Kleyna03, Wilkinson04, Munoz05} have 
shown a relatively flat velocity dispersion profile of $\sigma \approx 8-12 \kms$.
\citet{Kleyna03} (K03) used a two component model to test whether 
the second peak in photometry was a real feature.
They found a second kinematically distinct population 
with $\sigma = 0.5 \kms$ and $\Delta \ov = -1 \kms$. Our results lends support to this discovery by K03 but we do not agree on the magnitude of the velocity dispersion of the substructure. We discuss this in greater detail later. 

K03 argued through numerical simulations that the stellar clump they
discovered could survive if the dark matter halo of Ursa Minor had a large 
core (about $0.85~\rm kpc$)  but not a cusp like the prediction for inner parts of halos of $1/r$ from CDM simulations \citep{Navarro97}. 
Similar numerical simulations including the Ursa Minor stellar clump have
confirmed this result \citep{Lora2012}. Similar conclusions have been reached 
using the observed projected spatial distribution of the five globular clusters in Fornax dSph \citep{Mackey03}.
The survival of these old globular clusters has been interpreted as evidence that the dark
matter halo of Fornax may have a large core in stark contrast to the predictions of dark-matter-only CDM simulations  \citep{Goerdt06, Sanchez2006,
Cowsik09, Cole2012}.
Thus, the study of the properties of the substructure in Ursa Minor has far reaching implications 
for the dark matter halo of this dSph 
and by extension the properties of the dark matter particle. Our study is complementary to the recent studies using the presence of multiple stellar populations in Fornax and Sculptor that also seem to point towards a cored dark matter density profile \citep{Battaglia08, Walker2011,
Amorisco2012}. 

Current methods for finding kinematic substructure in the dSphs has relied 
on likelihood comparison parameter tests \citep{Kleyna03, Ural10}, non-parametric 
Nadaraya-Watson estimator \citep{Walker06}, or metalicity cuts and  
kinematics \citep{Battaglia11}, but not Bayesian methods. 
\cite{Hobson03} presented a Bayesian method for finding objects in noisy data.
The object detection method is able to find two or more objects using 
only a two component model in photometric data.
This method can be extended to include spectroscopic line-of-sight velocity data to search for objects using kinematics, as well as structural properties. 
We extend and apply this method to Ursa Minor to search 
for stellar substructure \citep{Irwin95, Kleyna98} and the kinematically cold  
feature found by K03.

\subsection{Data and Motivation for more Complex Models} 

\begin{figure*}
\includegraphics[width=83mm]{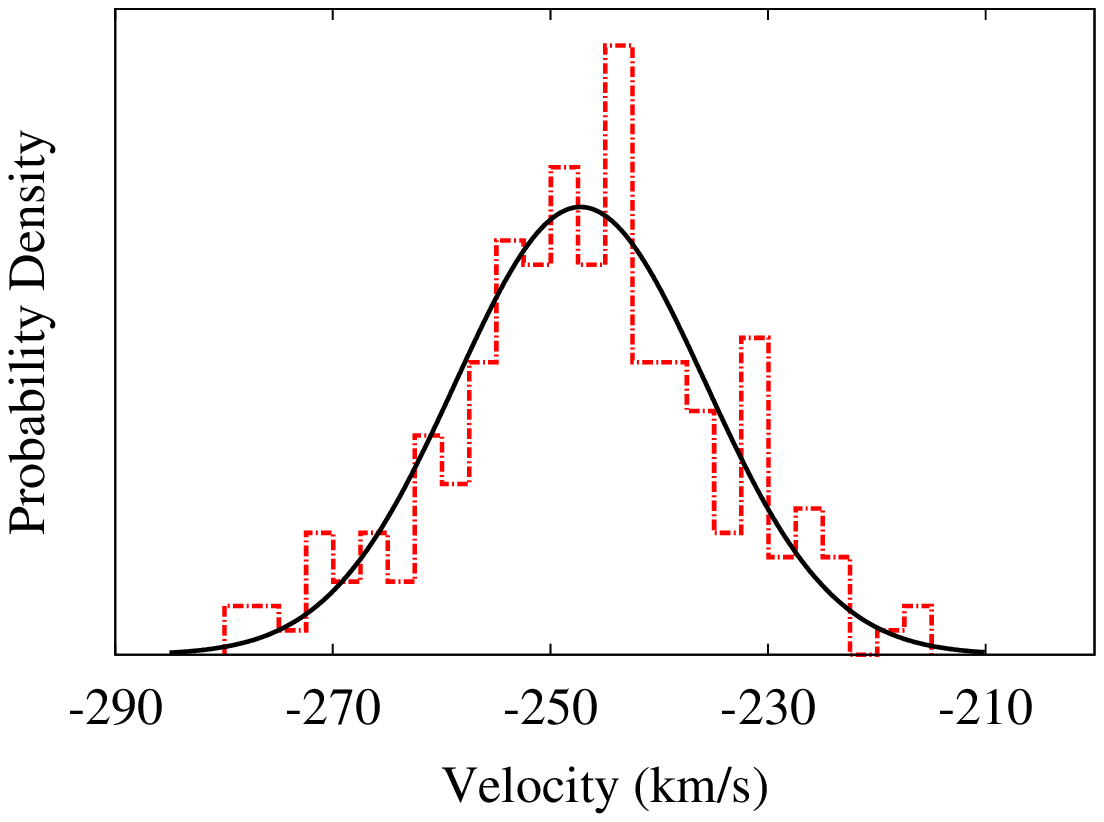}
\hspace{2mm}
\includegraphics[width=83mm]{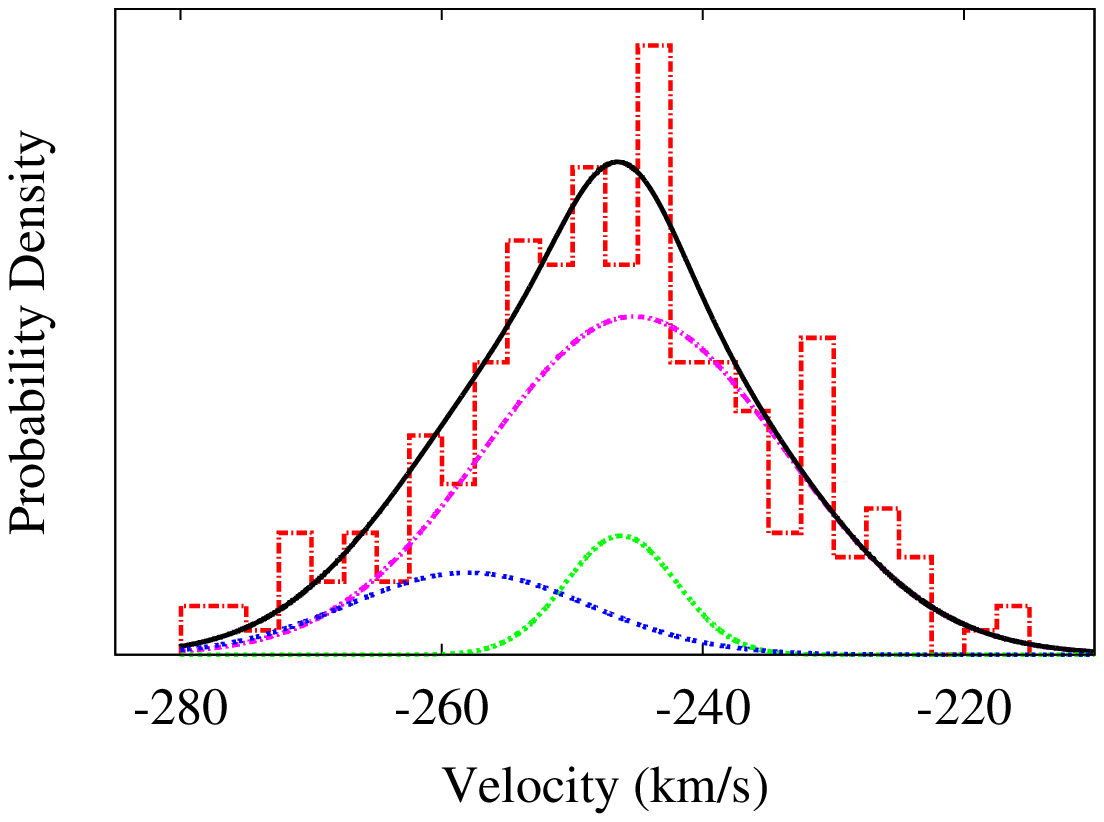}
\caption{ The binned line-of-sight velocity data (red dashed) in Ursa Minor.
{\em Right:} Over-plotted is the most probable Gaussian with $\sigma = 11.51$ and an $\ov = -247.25$ (black solid) from the null model (single Gaussian component).
{\em Left:} The line-of-sight velocity distributions of the secondary objects and primary populations.The lines correspond to the velocity dispersions of different populations found with the Bayesian object detection method; velocity offset object (blue dot-dot-space), cold object (green dotted), primary distribution (purple dot-dash), and the total (black solid). 
Each component is weighted by its average number of stars found using the Bayesian object detection method.
The additional kinematic components provide a better fit to the Ursa Minor data.
}
\label{fig:vel}
\end{figure*}

The spectroscopic data used contains 212 Ursa Minor member stars \citep{Munoz05}; the sample that K03 used to discover the cold feature contained 134 stars.
Figure ~\ref{fig:vel} (left) shows the radial velocities binned with the best fit single component Gaussian: this is a reasonable fit.
The data are, however, fit better if we use a three component Gaussian model, cf., Figure~\ref{fig:vel} (right). The mean and dispersion of these Gaussian distributions were derived from our Bayesian object detection that is the subject of this paper. As a prelude to our final results, we note that the centers of all three populations (the primary and two secondaries) found through the object detection method are spatially segregated.

\begin{figure*}
\includegraphics[width=175mm]{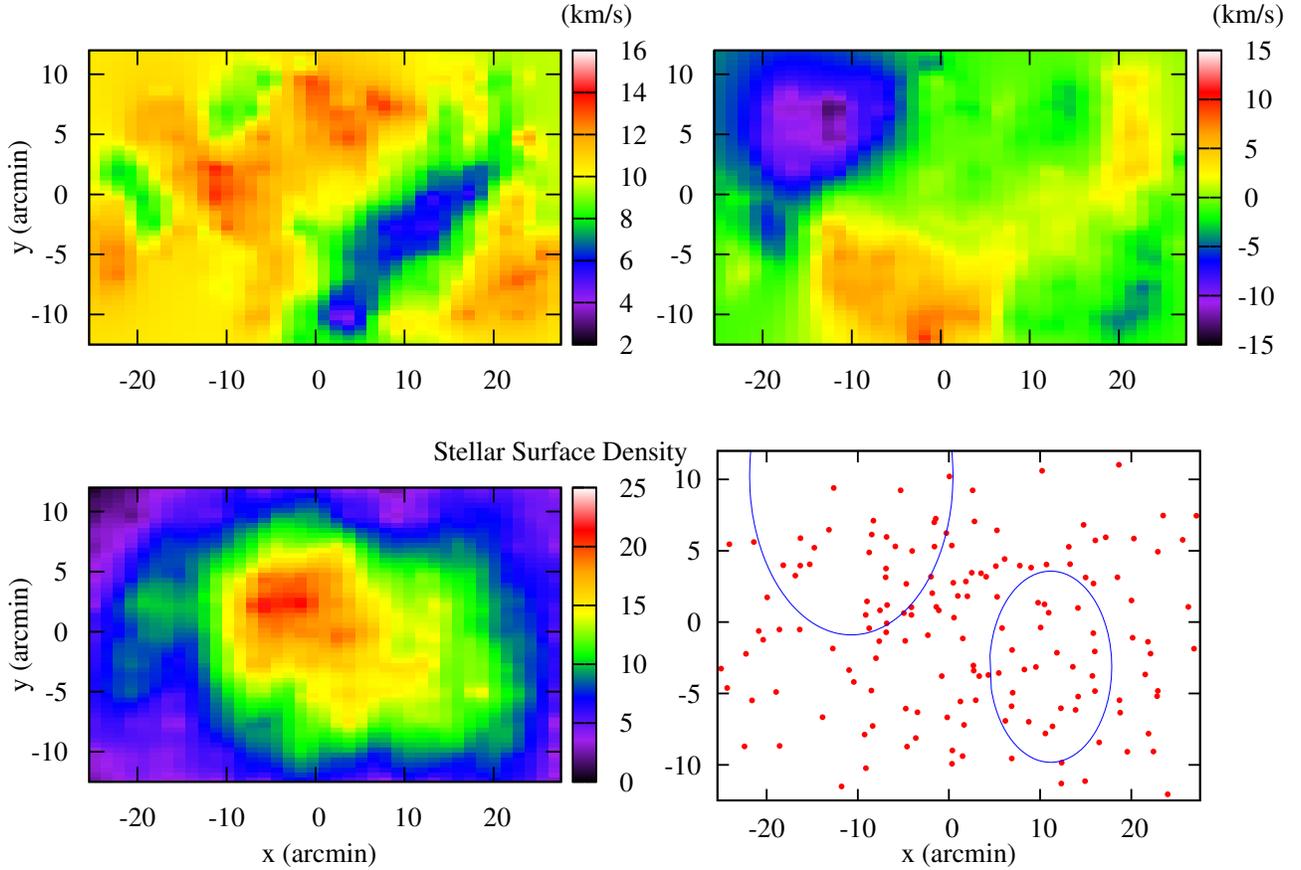}
\caption{The local kinematics of Ursa Minor using the \citet{Munoz05} data set.
{\em Upper Left:} A map of the velocity dispersion of Ursa Minor. A portion of the lower right quadrant drops below $6~\kms$ while the rest of the galaxy is relatively uniform.
{\em Upper Right:} The average velocity of Ursa Minor found concurrently with the velocity dispersion.  
In the upper left quadrant the deviation $\Delta \ov > 10-15~\kms$ relative to Ursa Minor while the rest of the galaxy does not differ more than $5~\kms$.
To make the contour plots, the velocity dispersion and the average velocity were found within a $5' \times 5'$ bin ($5' \simeq 110$ pc for a distance of 77 kpc).  
{\em Lower Left:} The stellar density profile of the stars in the \citet{Munoz05} data set. 
{\em Lower Right:} The most probable locations and sizes (tidal radii) of the two objects using the Bayesian object detection method in Ursa Minor.
Both of these locations correspond to the deviations seen in the average velocity and velocity dispersion maps.
The coordinate system used here is such that the x-axis lines up with the major axis which has a position angle of 49.4$^{\circ}$ \citep{Kleyna98}.
The adopted center for Ursa Minor was RA = $15^h 09^m 10^{s}.2$, DEC = $+67^{\circ}12'52"$ (J2000.0) (K03).
For the entire sample, we obtain a mean velocity $\ov =  -247.25~\kms$ and velocity dispersion $\sigma = 11.51~\kms$.  
}
\label{fig:contour}
\end{figure*}

Before we develop the Bayesian methodology, we would like to dissect the data to see if secondary populations are visible as strong local deviations in either mean velocity or velocity dispersion. To this end, we grid a $50' \times 30'$ region around the center of Ursa Minor finely and for each grid point, we find the average velocity $\ov$ and velocity dispersion $\sigma$ in a $5' \times 5'$ bin using the expectation-maximization (EM) method  (see Equations 12b and 13 of \citet{Walker09b}). We disregard grid points where there are fewer than 7 stars in the bin. We have plotted the smoothed $\sigma$ and $\ov$ maps created using this method in Figure~\ref{fig:contour}. The velocity dispersion map is the upper left panel and the average velocity map is the upper right panel.
The data is rotated such that the major axis is aligned with the abscissa  ($\theta = 49.4^{\circ}$, see Table 1 for the photometric properties of Ursa Minor we use).
There are two interesting features evident: in the $\sigma$ map, roughly centered at $(11', -4')$, $\sigma$ is significantly lower than the rest of the galaxy ($\sigma < 6 \kms$), and in the $\ov$ map centered at $(-13', 6')$, the $\ov$ significantly differs from Ursa Minor's overall average ($\Delta |\ov| > 10 \: \kms$).
For reference, the entire data set has $\sigma = 11.5 \: \kms$ and $\ov = -247.2 \: \kms$ with the EM method and $\sigma = 11.6 \pm 0.6 \: \kms$ and $\ov = -247.2 \pm 0.8 \: \kms$ using a single component Gaussian model sampled with a Bayesian nested sampling technique (see next section for an explanation of the Bayesian methods we use).
We have also plotted the number density (lower left panel) and the positions of the stars (lower right panel) in Figure~\ref{fig:contour} to provide a sense for where the data is and how significant the features in the $\ov$ and $\sigma$ maps are. 
The number density map is created the same way as the $\ov$ and $\sigma$ maps and it shows that both features are in regions that are reasonably sampled.
In the plot with the positions of the stars, we have also indicated the most probable locations for the centers and the extent of the the two features as found by our Bayesian object detection method.  We caution the reader that the plotted extents (tidal radii) of the these features have large error bars see Table~\ref{tab:prior}).

The center of the dip in the velocity dispersion (upper left panel of Figure~\ref{fig:contour})  is near the spectroscopic feature found by K03 and the secondary density peak seen in the photometry by several authors \citep{Irwin95, Kleyna98, Bellazzini02, Palma03}.
The average velocity feature we see does not correspond to any previous noted photometry or kinematic features. However, we note that the stellar isodensity contours of Ursa Minor are significantly asymmetric \citep{Kleyna98, Palma03} and could hide both features.

Here we aim to show that these two localized kinematic features in Ursa Minor are statistically significant.
We now turn to describing our Bayesian object detection method for finding secondary objects and model selection methods for assessing their significance.

\section{Methodology:  Theory}

This paper has two primary objectives:  to present a statistical methodology for detecting discrete features within a kinematic data set and apply this methodology to the Milky Way satellite galaxy Ursa Minor.  In this section we detail the statistical techniques used to detect kinematic objects within the Ursa Minor data set.
The pertinent question we are addressing is whether statistically distinct kinematic objects can be detected within a galaxy's stellar line-of-sight kinematic data and, if such an object is detected, how certain can we be that this object is an actual physical attribute of the system. Thus we require that any methodology used to detect multiple smaller composite objects within the kinematic data set have two important properties. 
First, any proposed algorithm must be able to discern an unspecified numbers of statistically separable features within the a galaxy's kinematic data set. And second, this methodology must allow for some kind of determination of the significance of a proposed object detection. 

To meet these criteria, we employ a Bayesian object detection technique first introduced by \citet{Hobson03}.
In our implementation, the data distribution is modeled with two separate components:  a background distribution referred to as the primary distribution, in our case, the Ursa Minor dSph ($\mathcal{P}_p$), and a 'secondary' distribution ($\mathcal{P}_s$) which is interpreted here as a feature or object of the Ursa Minor data set.  
Thus, the actual distribution is of the form:
\begin{equation}
\label{eq:pm-stuff}
 \mathcal{P}(d_i | \mathscr{M}) = (1-F) \mathcal{P}_p(d_i | \mathscr{M}_p) + F \mathcal{P}_s(d_i | \mathscr{M}_s)
\end{equation}
where $F$ is the total fraction of stars in the secondary population, $d_i$ represents an individual element of the Usra Minor data set $\mathscr{D}$ ($\mathscr{D} = \{d_i\}$), 
and $\mathscr{M}$ denotes the parameter set of the respective distribution's model.  A major benefit of this type of analysis is that data sets with multiple features will cause the secondary population parameter posteriors to become multi-modal where each individual mode represents a unique feature.  This enables us to search for an arbitrary number of objects without requiring an overly complicated probability distribution.  In addition, the local Bayesian evidences of each mode can be used as a selection criterion.
The evidence $Z \equiv \mathcal{P}(\mathscr{D} | H)$ 
is equal to the integral of the product of the likelihood, $\mathcal{L}(\mathscr{M}) \equiv \mathcal{P}(\mathscr{D} | \mathscr{M}, H) = \prod_i \mathcal{P}(d_i | \mathscr{M}, H)$, 
and prior probability, $Pr(\mathscr{M}) \equiv \mathcal{P}(\mathscr{M} | H)$:
\begin{equation}
 Z = \int \mathcal{L}(\mathscr{M}) Pr(\mathscr{M}) d\mathscr{M}\enspace.
\end{equation}
Here, the probability density of the parameter set $\mathscr{M}$ ({\em i.e.}, $\mathcal{P}(\mathscr{M} | \mathscr{D}, H)$), or posterior, is related to the evidence by the Bayes' theorem
\beq
\mathcal{P}(\mathscr{M} | \mathscr{D}, H) = \frac{\mathcal{P}(\mathscr{D} | \mathscr{M} , H) Pr(\mathscr{M})}{Z} \enspace,
\label{eq:bayeseq}
\eeq
Later, we use the evidence as a criterion for selecting between two models, or hypotheses ($H$):  One that assumes a `secondary' feature
represented by equation \ref{eq:pm-stuff} ($H_1$) and another `null hypothesis' that only assumes the background distribution $\mathcal{P}_p$ ($H_0$).
In section \ref{sec:crit} we use this both directly in the ratio of evidences, or Bayes factor, and indirectly in the determination
of the the Kullback-Leibler divergence, a quantity the quantifies the amount of information gained from the assumption of one hypothesis over another.
Through a large set of Monte Carlo simulations, these criteria are then used to derive confidence levels on the exclusion of the null hypothesis.

Calculation of the above quantities and sampling of the posterior space was done utilizing a Bayesian nested sampling 
technique \citep{Skilling04, Feroz09}.  The reason for this choice
is that this sampling algorithm possesses all the capabilities required for this project: 
multi-modal posteriors can be explored efficiently, and the evidence is inherently evaluated.

\subsection{Likelihood}

Our methodology utilizes a two component probability distribution similar to that in the K03 paper (also see \citet{Martinez11}).
We base the `primary' ($p$) and `secondary' ($s$) probability distributions on a Gaussian with mean velocity 
$\ov_{p,s}$, using the velocity errors $\epsilon_i$, and the assumption of a constant 
velocity dispersion, $\sigma_{p,s}$, as the spread:
\beq
	\mathcal{P}_{p, s}(v_i, R_i | \mathscr{M}_{p, s}) = \frac{\exp \left[ -\frac{1}{2}\frac{(v_{i} - \overline{v}_{p, s})^{2}}{(\sigma_{p, s}^2 + \epsilon_i^2)}\right]}{\sqrt{2 \pi (\sigma_{p, s}^2 + \epsilon_i^2)}} \frac{\rho_{p, s}(R_i)}{N_{p, s}} \\
\label{eq:partlike}
\eeq
Here, $\rho_{p,s}(R)$ is the 2-d stellar number density normalized to the total number in the population ($N_{p, s}$).

Unfortunately, because of spatial selection biases, $\rho_{p,s}(R)$ is difficult to model.  To account for this uncertainty, 
we consider only the `conditional' likelihood (see \citet{Martinez11} for details):
\begin{equation}
 \mathcal{P}_{p,s}(v_i | R_i, \mathscr{M}) = \mathcal{P}_{p,s}(v_i, R_i | \mathscr{M})/(\rho_{p,s}(R_i)/N_{p,s}).
\end{equation}
With this, equation \ref{eq:pm-stuff} becomes:
\begin{equation}
\label{eq:pm}
 \mathcal{P}(v_i | R_i, \mathscr{M}) = (1-f(R_i)) \mathcal{P}_p(v_i | R_i, \mathscr{M}_p) + f(R_i)\mathcal{P}_s(v_i | R_i, \mathscr{M}_s)
\end{equation}
where $f(R_i)$ is now the `local' fraction of stars in the secondary population defined by
\begin{equation}
f(R_i)=  \frac{\rho_s(R_i | \mathscr{M}_s)}{\rho_s(R_i | \mathscr{M}_s) + \alpha \rho_p(R_i | \mathscr{M}_p)}
\label{eq:fraction}
\end{equation}
Here, we have introduced the variable $\alpha = N_s / N_p$.
Instead of varying $\alpha$ directly, we found that, in some instances, using total fraction as a free parameter simplifies the analysis:
\beq
	F_{total} = \frac{ \int \! \rho_{s} \, \mathrm{d}x\mathrm{d}y}{ \int \! \rho_{s} \, \mathrm{d}x\mathrm{d}y + \alpha \int \! \rho_{p} \, \mathrm{d}x\mathrm{d}y}.
	\label{eq:total_fac}
\eeq

For the primary population, we assume a king 2-d density profile whose parameters are fixed to the observed photometry.
The secondary object's density profile is taken to be a top-hat\footnote{Other profiles were tried including a King, and Plummer profile. We detected both  objects in all cases.  The scale radii for the stellar profiles were unconstrained and errors were higher in other cases.}.
Our Bayesian object detection model constituted of 8 parameters: 2 primary kinematic parameters, 2 secondary kinematic parameters, 
the x and y center and  tidal radius for the secondary population and the total fraction.
The  parameters, priors, and posteriors are listed in the first row of Table~\ref{tab:prior}.

\begin{table*} 
\label{tab:prior}
\begin{tabular}{|l|cc|ccc|}
\hline \hline
Parameter & Type & Prior (Units) &  Cold Spot & Velocity Offset \\
\hline \hline
\multicolumn{5}{c}{Model parameters from Bayesian object detection method}\\
\hline 
$\sigma_s$ & flat & Cuts 1/2 (see caption) & $3.5^{+1.75}_{-2.25}$ & $8.75^{+1.5}_{-2.25}$ \\
$\sigma_p$ & flat & 0 to 20 $\kms$ & $11.75 \pm 0.5$ & $10.75 \pm 0.5$  \\
$\ov_s$ & flat & Cuts 1/2 (see caption)   & $-246.75^{+1.75}_{-2.0}$ & $-258.75^{+2.0}_{-1.75}$ \\
$\ov_p$ & flat & -242 to -252  $\kms$ & $-247.5 \pm 0.75 $ & $-245.25 \pm 0.75$ \\
$x_{cen}$ & flat & -0.6 to 0.6 kpc & $0.25^{+0.04}_{-0.06}$ & $-0.24 \pm 0.09$  \\
$y_{cen}$ & flat & -0.4 to 0.4 kpc & $-0.07^{+0.03}_{-0.07}$ & $0.23 \pm 0.02$ \\
$r_{tidal}$ & flat in $\log_{10}$ & 10 to 300 (pc) & $151^{+53}_{-28}$ & $251^{+24}_{-22}$ \\
$F_{total}$ & flat in $\log_{10}$ & $10^{-5}$ to 1 & $0.79^{+0.21}_{-0.16}$ & $0.32^{+0.47}_{-0.26}$  \\
\hline
\multicolumn{5}{c}{Secondary Population Model Parameters from simultaneous 3-component modeling}\\
\hline
$x_{cen}$ & flat & $-0.24 \pm 0.1$ kpc & $0.26 \pm 0.02$ &  $-0.23^{+0.095}_{-0.035}$\\
$y_{cen}$ & flat & $0.23 \pm 0.1$ kpc & $-0.07 \pm 0.01$ &  $0.22 \pm 0.02$ \\
$r_{tidal}$ & flat in $\log_{10}$ & 10 to 300 pc & $151^{+151}_{-10}$ &  $269^{+26}_{-24}$\\
$\sigma_s$ & flat & Cuts 1/2 (see caption) & $4.25 \pm 0.75$ &  $9.25 \pm 1.25$\\
$\sigma_p$ & flat & 0 to 20 $\kms$ &$11.5 \pm 0.5$ & $11.5 \pm 0.5$  \\
$\ov_s$ & flat & Cuts 1/2 (see caption) & $-246.25 \pm 1.0$ &  $-258.0 \pm 1.5$ \\
$\ov_p $ & flat & -252 to -242 $\kms$ & $-245.25^{+0.75}_{-0.5}$ & $-245.25^{+0.75}_{-0.5}$  \\
$f_{local}$ & derived & -- &  70\% (15.8/22.5) & 85 \% (27.0/31.6)\\
\hline 
\hline
\end{tabular} \\
\caption{Parameters, Priors, and Posteriors.
$\sigma_s$ and $\sigma_p$ are the velocity dispersions of the secondary and primary populations.  $\ov_s$ and $\ov_p$ are the average velocities of the secondary and primary populations.  $x_{cen}$ and $y_{cen}$ refer to the $x$ and $y$ centers of the secondary population.  Note that the data was rotated such that the x axis and the major axis are parallel.  $r_{tidal}$ is the tidal radius in a top hat model for the secondary population.  $F_{total}$ is the ratio of stars in the secondary population to the total population.  For the first section, the 4th and 5th columns denote the values when detecting the two objects individually.  
The two cuts indicated in the table as ``Cuts 1 and 2" are defined as follows. Cut 1 is $0 \leq \sigma \leq 10 \kms$ and $-252 \leq \ov \leq -242 \kms$ to find the cold spot object.  Cut 2 is $0 \leq \sigma \leq 20 \kms$ and $-267 \leq \ov \leq -237 \kms$ to find the velocity offset object. 
In the second section, the 4th and 5th column denote the values calculated for the two objects simultaneously using a 3-component model.  The coordinates $x_{cen}$ and $y_{cen}$ of the objects were only allowed to vary within $\pm 0.1 \rm kpc$ of the value obtained from the Bayesian object detection method. 
$f_{local}$ is the weighted average fraction of secondary population stars in each secondary object's location.}
\end{table*}

\subsection{Model Selection}
\label{sec:crit}

Even with accurate probability density modeling and thorough parameter space exploration, 
any object detection methodology will have fairly limited capabilities if the significance of a detection cannot be determined.
In our method, we use several model selection techniques 
to assess the significance of finding such an object.
Here, the posterior, likelihood, and evidence are used as the basis for determining selection criteria 
that measure the suitability of an hypotheses.  The two hypotheses that are compared are a model that contains
no sub-component feature (the `null hypotheses' ($H_0$)) and a model containing a sub-population ($H_1$).
Model selection techniques generally fall into two categories: those derived from the Bayesian evidence, 
and those based on information theory (specifically the Kullback-Leibler divergence ($D_{KL}$) or information entropy).
Among the most common are the Bayes factor, the Bayesian information criterion (BIC), 
the Akaike information criterion (AIC) \citep{Akaike74}, the Deviance information criterion (DIC) \citep{Spiegelhalter2002}, and direct calculation of the Kullback-Leibler divergence  ($D_{KL}$) \citep{Kullback1951}.
(For a review and the use of information criterion in cosmology see \citet{Liddle07}, for more general reviews of of model selection particularly Bayesian methods in cosmology see \citet{Liddle06, Trotta08}.)   
In this paper we use the Bayes Factor, DIC, and $D_{KL}$ to quantitatively derive confidence levels. We do not discuss the AIC or BIC since they are Gaussian approximations of the evidence and $D_{KL}$ respectively.

The Bayes factor is the ratio of the evidence of two models or hypotheses.
For example, the Bayes factor between two hypotheses $H_0$ and $H_1$, or single component versus multiple components is defined to be

\beq
	B_{01} = \frac{\mathcal{P}(\mathscr{D} | H_1)}{\mathcal{P}(\mathscr{D} | H_0)}.\\
\label{eq:bayesfactor}
\eeq
The general rule of thumb is that $B_{01} > 1$ favors hypothesis $H_1$ and $B_{01} < 1$ favors hypothesis $H_0$.
The significance of $B_{01}$ is usually computed as $\ln{B_{01}}$ with $\ln{B_{01}}< 1$, $1<\ln{B_{01}} < 2.5$, $2.5<\ln{B_{01}} < 5, \ln{B_{01}} > 5$ 
corresponding to inconclusive, weak, moderate and strong evidence, respectively, in favor of hypothesis $H_1$.  
The Bayes factor has the advantage that it is an output of our sampling algorithm.  But, the main disadvantage is that 
the Bayes factor inherently penalizes the model whose parameter space has the larger degrees of freedom. 
This can make determination of the significance of a detection ambiguous in that the Bayes factor
will naturally underestimate the importance of a proposed detection.  We address this issue by
first utilizing additional selection criteria based on information theory and second, 
null hypothesis mock data set analyses.

As mentioned in the previous paragraph, we wish to supplement the Bayes factor with other
selection criteria based on information theory.  Typically, these criteria are derived from $D_{KL}$ that 
quantifies how much more information you gain by switching from one probability distribution to another.
For our case, this quantity is: 
\begin{equation}
 D_{KL}(\mathcal{P}_1,\mathcal{P}_0) = \int \ln\left(\frac{\mathcal{P}(\mathscr{M}|\mathscr{D}, H_1)}{\mathcal{P}(\mathscr{M}|\mathscr{D}, H_0)}\right) \mathcal{P}(\mathscr{M}|\mathscr{D}, H_1) d\mathscr{M}\enspace,
\end{equation}
where $\mathcal{P}_0, \mathcal{P}_1$ are the posteriors under hypotheses $H_0$ and $H_1$, respectively.
Another quantity, the DIC \citep{Spiegelhalter2002}, is related to the amount of information gained through the full posterior as opposed to assuming only the prior probability distribution (i.e., $D_{KL}(\mathcal{P}, \mathrm{Pr})$):
\begin{equation}
 \mathrm{DIC} \equiv -2\widehat{D_{KL}}(\mathcal{P}, \mathrm{Pr}) + 2\mathcal{C}_b
\end{equation}
where $\mathcal{C}_b \equiv \overline{\chi^2(\mathscr{M})} - \chi^2(\overline{\mathscr{M}})$, $\chi^2 \equiv -2\ln(\mathcal{L})$, 
and $\widehat{D_{KL}}(\mathcal{P}, \mathrm{Pr}) \equiv \ln(\mathcal{L}(\overline{\mathscr{M}})) - \ln(Z)$ \citep{Trotta08}.
We emphasize that the evidence or Bayes factor and $D_{KL}$ should be used over the traditional information criterion whenever possible.
We also introduce the total membership as a physically interpretable model selection method tailored for the  problem at hand.
The membership that a star is part of the secondary population is derived from the posterior by the ratio of the secondary likelihood to total likelihood \citep{Martinez11}.
For the ith star, the membership is:
\beq
	m_{i} = \frac{f(R_i) \mathcal{P}_s(v_i | R_i, \mathscr{M}_s)}{(1-f(R_i)) \mathcal{P}_p(v_i | R_i, \mathscr{M}_p) + f(R_i) \mathcal{P}_s(v_i | R_i, \mathscr{M}_s)} \\
\label{eq:member}
\eeq
\noindent As the membership is derived from the posterior, each star will have its own probability distribution.
Our data set contains 212 stars and so to simplify the analysis we use the average membership of each star's probability distribution.
A global model selection parameter, the total average membership, can be found and interpreted as the average number of stars contained in the secondary population.
We find (see Figure~\ref{fig:dkl-mock1}-\ref{fig:dkl-mock2}) that the membership correlates with each of the other model selection parameters (i.e., a model with high evidence will have high membership and a model with low evidence will have low membership).

\subsection{Testing the Method with Mock Data}

\begin{figure*}
\includegraphics[width=175mm]{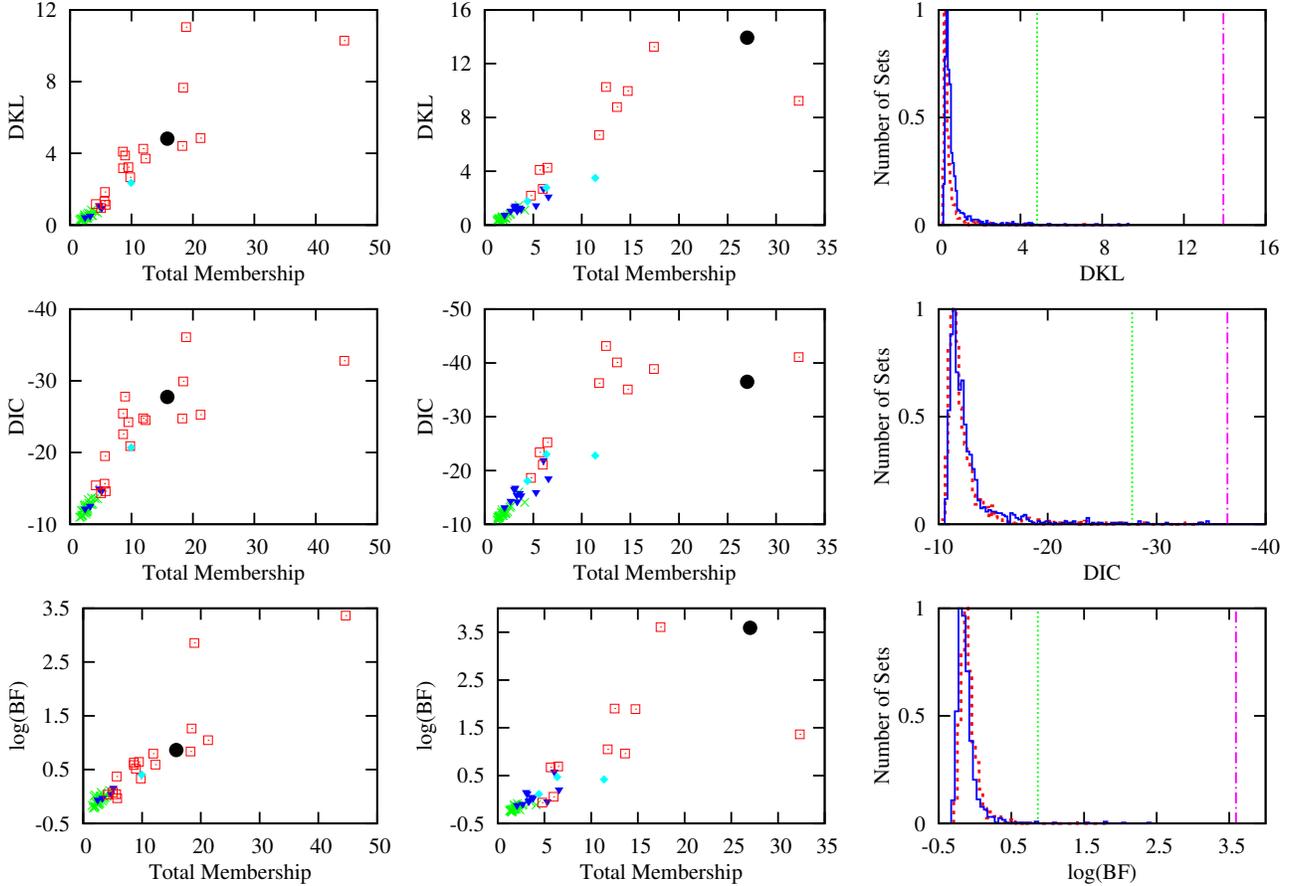}
\caption{Model selection tests using $D_{KL}$, DIC, $\log{BF} = \ln{\mathrm{B_{01}}}$ (cf., \S\ref{sec:crit} for definitions) for 50 mock data sets located at (0.2, -0.1). Also shown for comparison are the results for the actual Ursa Minor data set. A more negative DIC  favors the secondary object hypothesis more strongly, while the same is true for larger values of $D_{KL}$ and Bayes factor. 
{\bf Left column}: Figures in column 1 show the results of the analysis of the mock data sets in exactly the same way as the real data set was analyzed to look for the cold object with cuts on mean velocity and dispersion given by $0 \leq \sigma \leq 10 \kms$ and $-252 \leq \ov \leq -242 \kms$ (Cut 1). The top panel shows $D_{KL}$, the middle panel DIC and the bottom panel the logarithm of the Bayes factor (written in the text as $\ln{B_{01}}$. Mock data sets that had second populations with significant differences in their kinematics with respect to the background population were found with our object detection method. 
The symbols are labeled/colored according to a by-eye classification of the x and y posterior: peaked/found (red square), not peaked/ not found (green x), possible peaks (blue triangle) and double peaked with one correct (light blue diamond). The results for the actual Ursa Minor data set is shown as filled black circle. 
{\bf Middle column}: This panel has the same symbols and colors as the left most column. The difference here is that the velocity cuts used are broader (and the same as that used to find the velocity offset object). The cuts are $0 \leq \sigma \leq 20 \kms$ and $-267 \leq \ov \leq -237 \kms$ (Cut 2).
{\bf Right column}: Histograms of $D_{KL}$, DIC and Bayes factor from analyses of 1000 null hypothesis mock data sets with Cut 1 (red dotted) and Cut 2 (blue solid). 
The vertical lines show the $D_{KL}$, DIC and Bayes factor values (in the top, middle and bottom panels, respectively) for the actual Ursa Minor data set with Cut 1 (green dotted) and Cut 2 (magenta dot-dashed). 
}
\label{fig:dkl-mock1}
\end{figure*} 

\begin{figure*}
\includegraphics[width=175mm]{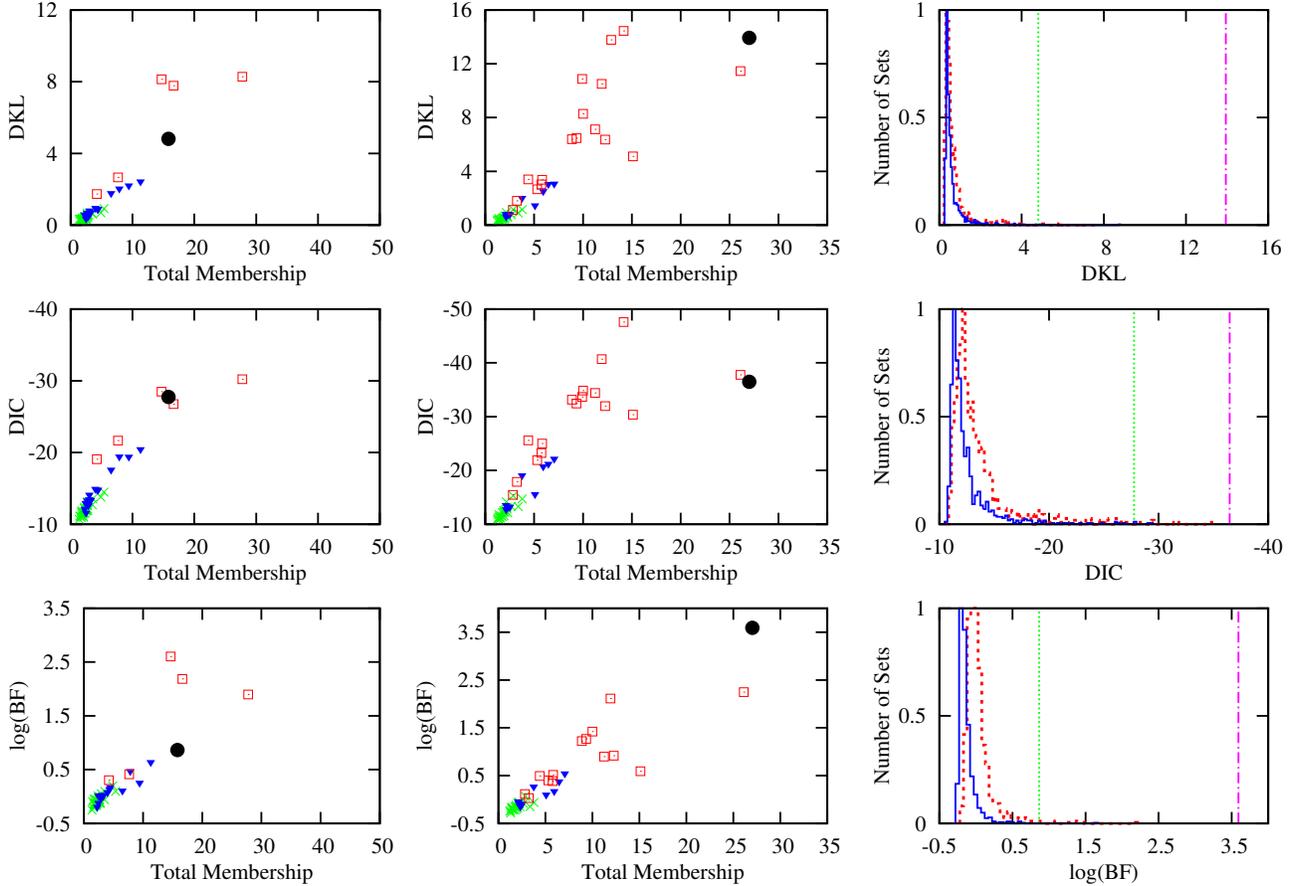} 
\caption{ Model selection tests using $D_{KL}$, DIC, $\ln{B_{01}}$ for 50 mock sets located at (-0.24, 0.23) and the Ursa Minor data. The layout is the same as Figure~\ref{fig:dkl-mock1}. The third column from left displays the results from the scrambled mock data sets instead of the null hypothesis mock data sets plotted in Figure \ref{fig:dkl-mock1}. 
}
\label{fig:dkl-mock2}
\end{figure*} 

\begin{figure*}
\includegraphics[width=175mm]{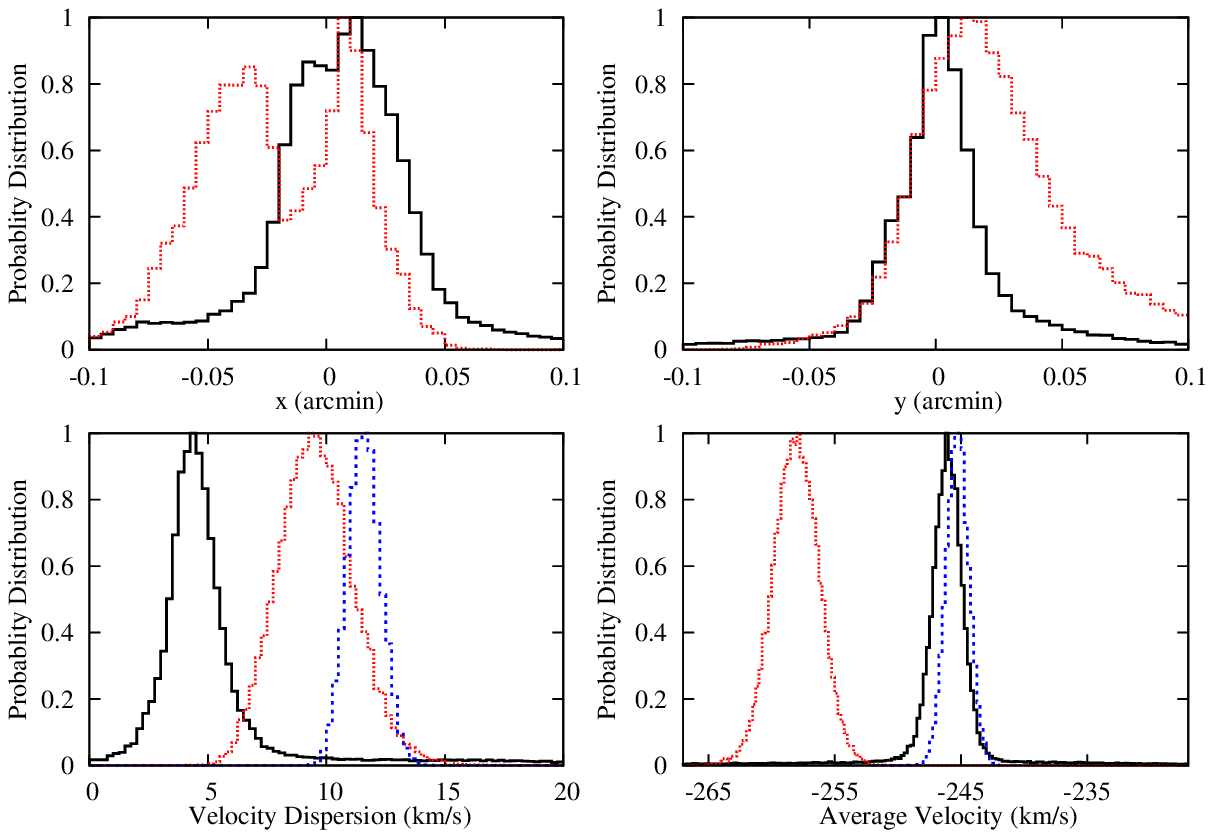}
\caption{The posteriors for the secondary populations in Ursa Minor using the three-parameter model.
The secondary populations are fixed at $(0.25, -0.07)\rm kpc$ and $(+0.24, 0.23)\rm kpc$ and allowed to vary $0.1 \rm kpc$ in both coordinates. They correspond to the cold (black solid) and the velocity offset (red dots) objects, respectively. 
{\em Upper Left:}  The x coordinate posteriors for of the secondary populations.
{\em Upper Right:} The y coordinate posteriors for the secondary populations.
{\em Lower Left:} The velocity dispersion posteriors of the cold object (black solid), velocity offset object (red dotted), and the primary (blue dashed). 
{\em Lower Right:} The average velocity posteriors of the cold object (black solid), velocity offset object (red dotted), and the primary (blue dashed).
The secondary populations have distinct kinematic properties and are both localized.
}
\label{fig:posterior_more}
\end{figure*}

We created 100 mock data sets containing a second population to test whether known secondary objects could be detected using our object detection method.
The second populations were located at either (0.2, -0.1) or (-0.23, 0.24) kpc (roughly the locations of the cold and velocity offset objects).
The kinematic and structural parameters of this second population were selected to mimic the cold and velocity offset objects.
The positions and velocity errors from the Ursa Minor data set were used to simulate observational errors.
To pick which population a star is assigned to, the local fraction was found via Equation~\ref{eq:total_fac} and membership was randomly assigned with the second population weighted by the local fraction.
The primary population parameters were the best fit values from Ursa Minor photometry and the kinematics of the entire sample: $r_{tidal} = 1.745\ \rm kpc$, $r_{core}=0.401\ \rm kpc$, ellipticity $\epsilon_p =0.56$, $\sigma = 11.5~\kms$, and $\ov = -247~\kms$.
The second population's base parameters were: $\epsilon_s = 0$, $\theta_s = 0.0$, $F_{total} = 60/212$, $r_{core}= 0.05 \rm kpc$, $\Delta \ov_s = 0~\kms$, $\sigma = 4 \kms$, $r_{tidal} = 0.15 \rm kpc$ for (0.2, -0.1) location. For the (-0.23, 0.24) location, we used a slightly larger value for tidal radius, $r_{tidal} = 0.25 \rm kpc$. We note that both populations were created assuming an underlying King profile but the object detection used a top-hat model when finding the second population, identically to how the objects were found in the actual data.
Each individual mock data set had 1-3 secondary parameters that deviated from the base parameters to test how each parameter effected the detection. In some sets we did not expect to find the secondary population, for example, if they had small tidal radius or small secondary population fraction.

The results for model selection of the $D_{KL}$, DIC, $\ln{B_{01}}$, and total membership using two different kinematic priors are summarized in the right and middle columns of Figure \ref{fig:dkl-mock1} (secondary population located at (0.2, -0.1)) and Figure \ref{fig:dkl-mock2} (secondary population located at (-0.23, 0.24)).
The left and middle columns show different kinematic priors with the left column showing the cuts to find kinematically cold objects ($0 \leq \sigma \leq 10 \kms$, $-252 \leq \ov \leq -242 \kms$). The middle has the cuts to find objects with a significant velocity offset ($0 \leq \sigma \leq 20 \kms$, $-267 \leq \ov \leq -237 \kms$); this cut will also find the kinematically cold objects, but in the Ursa Minor case the velocity offset object was significantly more likely and tended to dominate the posterior. 
The symbols for these columns are labeled/colored according to a by-eye definition of the x and y posterior: peaked/``found" (red square), not peaked/``not found" (green x),  ``possible" peaks (blue triangle), double peaked with one correct (light blue diamond). Results for the actual Ursa Minor data with corresponding cuts are shown as filled black circle.
The ``possible" peaks are posteriors where there was a peak near the second population's center, a small/medium peak somewhere else in the posterior, or a small peak at the correct location.
The double peaked data had one peak at the correct location and a second at another location.
The ``possible" sets tended to span the border between ``found" and ``not found" and were not easily  categorized otherwise.

Both Figures show a clear trend between the ``found" and ``not found" sets in all the model selection methods. Note that more negative DIC corresponds to favoring the more complicated model.
Sets that are ``not found" by-eye have model selection criteria that is equivalent to the model selection criteria of null hypothesis mock data sets (i.e., sets made with no second population), cf., Section~\ref{sec-null}.
The model selection criteria for the two objects found in Ursa Minor also lie in the ``found" section of the mock data's selection criteria. 
From the analysis of these mock data sets we conclude that our method is fully capable of detecting the cold and velocity offset objects, and the model selection criteria favor the favor presence of two additional components in Ursa Minor.

\section{Results}
\label{section:results}

We have found two objects in the Ursa Minor data set of \citet{Munoz05} using a Bayesian object detection method.  The first object, referred to as the ``cold object" here, is kinematically cold, $\sigma_{cold} = 3.5^{+1.8}_{-2.3} \rm \kms$, with an average velocity close to that of the full Ursa Minor sample, $\ov_{cold} = -246.8^{+1.8}_{-2.0} \kms$ .  The location coincides 
with the location of the K03 stellar clump. The second object, referred to as the velocity offset object, has a large average velocity offset compared to the mean velocity of Ursa Minor,  $\ov_{vo} = -258.8^{+2.0}_{-1.8} \kms$ with a dispersion of $\sigma_{vo} = 8.8^{+1.5}_{-2.3} \kms$.
The kinematics and structural properties are summarized in the first section of Table~\ref{tab:prior}.
The model selection tests for the cold object are: Total Membership $= 15.8$, $D_{KL} = 4.8$,  $DIC = -26.1$, $\ln{B_{01}} = 0.9$. 
The model selection tests for the velocity offsets object are: Total Membership $= 27.0$, $D_{KL} = 13.9$,  $DIC = -36.5$, $\ln{B_{01}} = 3.6$.
In Figures~\ref{fig:dkl-mock1}-~\ref{fig:dkl-mock2} the results of model selection test are plotted along side the mock set distributions.
All of the model selections tests favor the additional secondary objects with moderate to high significance except for the Bayes factor which has weak to moderate significance for the cold and velocity offset objects. This significance is based on the recommendations of \citet{Trotta08,Ghosh2006, Spiegelhalter2002}. However, it is important to judge the significance of the information criteria and the Bayes factor for the problem at hand. We do this by generating mock data sets and deriving the information criteria and Bayes factor in the same way as the real data is handled. 
When this test is performed, we find that the confidence levels of both objects are above the 98$\%$ C. L. (see Table~\ref{tab:prior}). In addition, all of the model selection values, for both locations/objects, lie in the ``found" region of the mock sets of Figure~\ref{fig:dkl-mock1}-\ref{fig:dkl-mock2}.

\subsection{Significance of Information Criteria and Bayes' Factor}
\label{sec-null}

In order to assess the significance of the model selection tests, knowledge of the false positive rate is helpful.
We make use of two types of tests: null hypothesis mock data sets and scrambled data sets.
Null hypothesis mock data sets are constructed by redrawing the line-of-sight velocities from a Gaussian with Ursa Minor kinematics\footnote{We used $\ov = -247.0 \kms$ and $\sigma = 11.5 \kms$.}.
To simulate positional and velocity errors, the positions of stars and the line-of-sight velocity errors were kept.
The scrambled sets were constructed by repicking a random observed line-of-sight velocity and line-of-sight velocity error pair, without replacement, for each star in the data set.
1000 null hypothesis mock data sets and scrambled data sets were constructed and analyzed with our object detection method.

The results of the object detection method and our employed model selection tests for the null hypothesis mock data sets and the scrambled mock data sets are shown in the last columns of Figures~\ref{fig:dkl-mock1} and \ref{fig:dkl-mock2}, respectively. 
The $D_{KL}$ (top), DIC (middle), and $\ln{B_{01}}$ (bottom) are binned and the maximum is normalized to unity. The analysis with the cuts to find cold objects ($0 \leq \sigma \leq 10 \kms$, $-252 \leq \ov \leq -242 \kms$) is shown in red, while that with cuts to find objects with significant velocity offset ($0 \leq \sigma \leq 20 \kms$, $-267 \leq \ov \leq -237 \kms$) is shown in blue. 
The model selection results for the real Ursa Minor data are plotted as vertical lines: cold object with green dotted line and velocity offset object with purple dash-dot line. 
The confidence levels of the model selection criteria for the null hypothesis mock data sets and scrambled data sets are above the 98.5$\%$ c. l. with every model selection criteria.  
They are summarized in Table~\ref{tab:mocksets}.
Even though the $\ln{B_{01}}$ shows weak evidence for the cold object according to standard definitions, it is still above the $95 \%$ confidence level for both the null hypothesis mock data sets and scrambled data sets. 

\begin{table*}
\label{tab:mocksets} 
\begin{tabular}{|l| cccc|}
\multicolumn{5}{c}{\em Test using null hypothesis mock data sets} \\
\hline
\hline
& Total Average  & \multicolumn{2}{c} {Information Entropy} & Bayesian Evidence\\
& Membership & $D_{KL}$ & DIC & $\ln{B_{01}}$ \\
\hline
Value at 95\% C.L. from null hypothesis mock data sets using Cut 1 & 5.25 & 1.28 & -16.35 & 0.17 \\
Cold object values from data (inferred C. L.) & 15.82 (99.8\%)  & 4.82 (99.7\%) & -26.08 (99.5\%) &  0.87 (99.7\%)\\
Value at 95\% C.L. from null hypothesis mock data sets using Cut 2 & 4.49 & 1.84 & -17.79 & 0.13 \\
Velocity offset object values from data (inferred C.L.) & 27.02 ($>99.9$ \%) & 13.93 ($>99.9$ \%)& -36.49 (99.9 \%)& 3.59 ($>99.9$ \%)\\
\hline
\hline
\multicolumn{5}{c}{\ }\\
\multicolumn{5}{c}{\em Test using scrambled data sets} \\
\hline
\hline
& Total Average  & \multicolumn{2}{c} {Information Entropy} & Bayesian Evidence\\
& Membership & $D_{KL}$ & DIC & $\ln{B_{01}}$ \\
\hline
Value at 95\% C.L. from scrambled mock data sets using Cut 1&  6.99 & 2.22 & -20.45 & 0.40\\ 
Cold object values from data (inferred C. L.) & 15.82 (99.7\%)  & 4.82 (99.1\%) & -26.08 (98.5\%) &  0.87 (99.0\%)\\
Value at 95\% C.L. from scrambled mock data sets using Cut 2& 3.89 & 1.46 & -16.30 & 0.07 \\
Velocity offset object values from data (inferred C.L.) & 27.02 ($>99.9$ \%) & 13.93 ($>99.9$ \%)& -36.49 ($>99.9$ \%)& 3.59 ($>99.9$ \%)\\
\hline
\hline
\end{tabular}
\caption{Confidence Levels computed from null hypothesis and scrambled mock data sets. The inferred C.L.. refers to the number of null hypothesis mock data sets and scrambled data sets sets that have a model selection value lower than that of the actual Ursa Minor data. The $95\%$ C.L. value is defined such that 95\% of the null hypothesis or scrambled data sets have a value below this. 
Both additional populations found in the Ursa Minor data are above the $98 \%$ C.L. for all the model selection methods. The two cuts indicated in the table as ``Cuts 1 and 2" are defined as follows. Cut 1 is $0 \leq \sigma \leq 10 \kms$ and $-252 \leq \ov \leq -242 \kms$ used to find the cold spot object in the data.  Cut 2 is $0 \leq \sigma \leq 20 \kms$ and $-267 \leq \ov \leq -237 \kms$ used to find the velocity offset object in the data. 
}
\end{table*}

\subsection{Narrowing down secondary population parameters using a 3-component model\label{sec:3comp}}

To reliably calculate the kinematic properties of the secondary objects we introduce a model with two secondary populations.  The additional populations are only allowed to vary by $0.1\ \rm kpc$ in both $x$ and $y$ from the best-fit center locations found in the Bayesian object detection method for the cold and velocity offset objects.  
Equation~\ref{eq:fraction} is changed to include the third component and instead of the normalization parameter, $\alpha = \frac{N_p}{N_s}$, there are now two normalization parameters, $\alpha_2 = \frac{N_2}{N_1}$, and $\alpha_p = \frac{N_p}{N_1}$ where $N_1$ and $N_2$ denote the normalization of the first and second object.  
The results for the kinematic parameters are: $\sigma_{cold} = 4.3 \pm 0.8 \kms$, $\ov_{cold} = -246.3 \pm 1.0 \kms$, $\sigma_{vo} = 9.3 \pm 1.3 \kms$, and $\ov_{vo} = -258.0 \pm 1.5 \kms$, respectively. These values are in full agreement with the values obtained using the two-component (Bayesian object detection) method. 

The normalization ratios, as defined, are not easily interpreted. So we introduce a derived parameter, local fraction or $f_{local}$, that is defined as the weighted average of stars with memberships greater than 50\% in the secondary population compared to the total number of stars within the secondary object's tidal radius. In short, it is a measure of the fraction of secondary stars in each object's location.  We derive $f_{local, cold} = 15.8/22.5$ or 70\% and $f_{local, vo} = 27.0/31.6$ or 85\%. Clearly, we are able to find these objects only because they seem to have a high local fraction. 
The kinematics and structural properties of the secondary population model are summarized in the second section of Table~\ref{tab:prior}.
In upper left and right panels of Figure~\ref{fig:posterior_more}, we have plotted the posteriors for the x and y centers, respectively, for the cold (black solid) and velocity offset objects (red dotted).  The centers for the cold and velocity offset object are $(0.25, -0.07)\ \rm kpc$ and $(+0.24, 0.23)\ \rm kpc$ and the two panels show the deviation from the ``fixed'' centers.  The lower right (lower left) panel of Figure~\ref{fig:posterior_more} is the posterior of the $\sigma_s$ ($\ov_s$) for the cold (black solid), velocity offset objects (red dotted), and primary (blue dashed).

An increased prior volume for the centers and tidal radius in the 3-component model changes the posteriors for the structural parameters of the velocity offset object but does not changes its kinematics.
By only allowing more freedom in the location of the centers (200 pc versus 100 pc) the posteriors of both centers gain tails. 
An increase in the maximum tidal radius (in the prior) of the objects (500 pc from 300 pc) increases the size of the velocity offset object and moves its center roughly 150 pc away from the center of Ursa Minor while the same change introduces tails in the posterior of the cold object.
Given these results, it is fair to say that the the size and center of the secondary objects are not known with precision  and more data will help considerably. However, our conclusions regarding kinematics seem to be robust. 

\subsection{Perspective Motion}

Line-of-sight velocity measurements for the Milky Way satellites receive a small contribution from $x$ and $y$ direction velocities of the star (where $z$ is along the line-of-sight to the center of the galaxy), and this contribution increases with distance from the center \citep{Feast1961,Kaplinghat08}. A similar contribution could also arise due to solid-body rotation or some other physical mechanism (such as tides) that leads to a velocity gradient. Motivated by the large velocity-offset we found, we ask whether the this term changes our conclusions. The observed line-of-sight velocity of a star may be written as,
\beq
    v_{\rm los} = v_z -v_x x / D - v_y y / D \\
\label{eq:perspective}
\eeq
where $D$ is the distance to the galaxy and $(x,y)$ are the projected coordinates on the sky.  
This method has been applied to the dSph's Fornax, Sculptor, Sextans, and Carina and results agree with other methods \citep{Walker08}.
The proper motion we find assuming only a primary population with a constant velocity dispersion is $(\mu_{\alpha}, \mu_{\delta}) = (529 \pm 848 , -280 \pm 449) \: \mathrm{mas \: century^{-1}}$, which shows clearly that we are unable to constrain the proper motion of Ursa Minor using this effect.

Observations from the HST find a proper motion for Ursa Minor of $(\mu_{\alpha}, \mu_{\delta}) = (-50 \pm 17,  22 \pm 16) \: \mathrm{mas \: century^{-1}}$ \citep{Piatek2005}, which is an order of magnitude smaller (when comparing the mean) than the result we calculate.
If stars with high membership in the velocity offset object are weighted as not being in Ursa Minor the proper motion of this subset is $(\mu_{\alpha}, \mu_{\delta}) = (117 \pm 90 ,   163 \pm 127) \: \mathrm{mas \: century^{-1}}$.
Removing both secondary populations this way results in $(\mu_{\alpha}, \mu_{\delta}) = (-84 \pm 79,   -185 \pm 174) \: \mathrm{mas \: century^{-1}}$.
If we remove all the stars in these locations we find $(\mu_{\alpha}, \mu_{\delta}) = (-67 \pm 60,   -203 \pm 181) \: \mathrm{mas \: century^{-1}}$.
These comparisons provide clear proof that it is hard to estimate the tangential velocity with perspective motion if there are secondary populations in the data set.

To investigate this issue further we run a three-component model to try and pin down the two secondary components when including perspective motion. 
We add this effect into our likelihood function by changing the model velocity for all three components (cf., $\overline{v}_{p, s}$ in Equation~\ref{eq:partlike}) to $v_{\mathrm{los},i}$ given by Equation~\ref{eq:perspective} with $x_i$ and $y_i$ for each star measured from the center of Ursa Minor. Each component has its own $v_z$ but $v_x$ and $v_y$ are the same for all three components.  Note that the actual tangential velocity of the two secondary components is now implicitly tied to the $v_z$ value -- there is no hope of disentangling them given the small projection on the sky of the secondary components. 
We then impose the same constraints on the center as before (cf., \S\ref{sec:3comp}). We find results that are consistent with those we found in \S\ref{sec:3comp} in the absence of perspective motion: $x_{cold} = 0.245^{+0.03}_{-0.04} \rm kpc$, $y_{cold} = -0.065^{+0.015}_{-0.025} \rm kpc$ and $x_{vo} = -0.275^{+0.04}_{-0.035} \rm kpc$, $y_{vo} = 0.24 \pm 0.025 \rm kpc$. The kinematic properties are the same as without perspective motion except the error bars are larger.  Thus the three-component model with the prior on the centers provides a different fit and favors the presence of the secondary objects over perspective motion. Had perspective motion or a velocity gradient or rotation been a better fit to the likelihood instead of either of the objects, this would not have been the case since the likelihood allows for the freedom to dial down the fraction of stars in the secondary objects. In this three-component fit, the mean velocity of Ursa Minor is $(-311 \pm 212,-548^{+357}_{-324},-245.5 \pm 0.75)~\kms$, in good agreement with the results obtained when stars in the locations populated by the secondary populations are removed.

Instead of using a three component model (as we did above), we also explored the effect of using the Bayesian object detection method including the perspective motion effect.  This could lead to faulty results (and we show below that it does) because the velocity offset spot has a large impact on the determination of the background parameters -- specifically the perspective motion.
With the velocity cuts to find the cold object, we find a mean velocity for Ursa Minor of $(-100^{+100}_{-100}, -1125^{+275}_{-250},-247.5^{+0.5}_{-0.5}) \kms$ and a dispersion in the line-of-sight velocity of  $11.0 \pm 0.5 \kms$. The dispersion of the cold object is now consistent with zero at about 1-$\sigma$, $3.25 \pm 3.0~\kms$ and the location of the centers is now much less well-determined. 
However, the values obtained for the perspective motion are unphysically large and hence this is clearly not the correct model to be considering.  
With the $\pm20 \kms$ velocity cut (to find the velocity offset object), we find  a mean velocity for Ursa Minor of $(-200^{+150}_{-150}, -1175^{+400}_{-400}, -247^{+1.0}_{-1.25}) \kms$ and $10.75 \pm 0.5~\kms$ for its dispersion in the line-of-sight velocity. The center, as with the other object, is no longer tightly constrained, and the hint for deviation in mean velocity for this object is muted ($-258^{+7.5}_{-4.5}~\kms$).
Thus, we arrive at the conclusion (unsurprisingly) that {\em varying background parameters} in Bayesian object detection methods can lead to faulty results in data sets containing multiple signals if those signals have a significant effect on the determination of the background parameters. 
In particular, for this analysis we saw that the presence of the velocity offset spot affects the magnitude and the direction of the inferred tangential motion and hence the object detection method has trouble fitting one secondary location and perspective motion.  But with two localized secondary populations and perspective motion the method still picks out both secondary objects. Thus, the three component model is preferred by this data set.

A tangential velocity measured using perspective motion could also be hiding a possible solid-body rotation. An order of magnitude estimate of this rotation speed would be $v_{rot} = \frac{R_{e}}{D} \sqrt{v_x^2 + v_y^2}\: $ ($R_e = 445 \pm 44 \: \mathrm{pc}, D = 77 \pm 4 \: \mathrm{kpc}$). 
Using the results presented in this section, we calculate: 
$v_{rot} \sim 7 \kms$ with entire data set, and $v_{rot} \sim 4 \kms$ when the velocity offset population is removed, and when both secondary populations are removed or when all stars near the secondary populations are removed.  
The rotation speeds are all comparable but in each estimate the rotation is about a different axis. 
The summary of our results from this section is that a larger data set is required to simultaneously constrain properties of the secondary populations and rotation or proper motion. The results of our three-component analysis suggest that the data prefer the presence of both secondary objects to perspective motion (or a rotation that masquerades as it). 

\section{Discussion}

K03 utilized a frequentist likelihood test with a two component kinematic model (Ursa Minor dSph plus a secondary population) similar to our Bayesian object detection method.
They discovered a stellar clump with a high likelihood ratio ($\sim 10^{4}$) located at ($10\arcmin$,$4\arcmin$) (on-sky frame) relative to the Ursa Minor center with parameters, $\sigma = 0.5~\kms$, $v_s = -1~\kms$ and clump fraction of 0.7 (fraction of stars in the second population).
The kinematically cold object found with our Bayesian object detection method is centered at $(10.8\arcmin \pm 1.8, 5.5\arcmin \pm 0.9)$ (on-sky frame relative to Ursa Minor center), has a size of $6.7\arcmin \pm 0.5$, with kinematic properties $\sigma = 4.25 \pm 0.75~\kms$, and $\Delta \ov = -1.1^{+1.5}_{-1.25}~\kms$.
The difference between our results and those of K03 lie in the velocity dispersion of the cold object. We have considerably more stars (in total roughly 212 to 134 of K03) and are therefore able to infer the dispersion with much greater confidence. We find the mean value for the velocity dispersion to be close to 4 $\kms$, similar to the dispersion of Segue 1 dSph \citep{Simon11}.

The main uncertainty in our estimates of the dispersion for cold and velocity offset objects is the presence of perspective motion or solid-body rotation. Perspective motion by itself cannot explain these secondary populations.
A three-component analysis (i.e., main Ursa Minor population and both secondary populations) with the coordinates of the centers fixed to within 0.1 kpc and including perspective motion (with unconstrained tangential velocity) prefers the presence of both the secondary populations. In this analysis, the velocity dispersions of the cold or velocity offset objects are not significantly different from the values obtained without including perspective motion.

To estimate the luminosity of the secondary objects, we use the total membership of the objects with the assumption that the stars were drawn uniformly from the three distributions of Ursa Minor.
We find the luminosity of the cold and velocity offset objects to be $4 \times 10^{4} \: L_{\odot}$ and $6 \times 10^{4} \: L_{\odot}$.
The luminosity of the K03 object is $1.5 \times 10^{4} \: L_{\odot}$, and given the uncertainties we would chalk this down as agreement between the two analyses.
The dynamical mass within half-light radius of dispersion supported systems can be estimated to about 20$\%$ accuracy using the line-of-sight velocity dispersions and the half-light radius \citep{Walker09c,Wolf10}. 
Assuming that the ratio of $r_{1/2}/r_{tidal}$ of the objects is the same as that of Ursa Minor, we find $M_{1/2} = 6 \times 10^{5} \: M_{\odot}$, and $M_{1/2} = 5 \times 10^{6} \: M_{\odot}$ for the cold and velocity offset object.
From this $M/L (r_{1/2}) \approx 30 \: M_{\odot}/ L_{\odot}$ and $M/L (r_{1/2}) \approx 175 \: M_{\odot}/ L_{\odot}$ for the cold and velocity offset objects.
If we use this same estimator to find the velocity dispersions assuming the objects are relaxed systems with only stellar components and $M/L=2$ (as in K03), we estimate a velocity dispersion of $\sigma = 1.0~\kms$ for both the cold and velocity offset objects. 
This differs from the velocity dispersion found through our object detection method by 4 $\sigma$ and 6.6 $\sigma$ for the cold and velocity offset objects, respectively.
Note that the estimator for $M_{1/2}$ assumes that the system is dynamical equilibrium, which may not be the case here. 
If our current results hold up with the addition of more data, then either these objects have highly inflated velocity dispersions due to the influence of motion in binary stellar systems or tidal disruption, or these objects really do have a much larger mass than inferred from their luminosities. In the latter case, we would have found a satellite of Ursa Minor, the first detection of a satellite of a satellite galaxy.  We discuss each of these possibilities briefly below.

Contribution of binary orbital motion to the line-of-sight velocities can inflate the observed line-of-sight velocities of stars \citep{Aaronson1987, Hargreaves1996, Olszewski1996, Minor10, McConnachie2010}. 
A galaxy with a lower intrinsic velocity dispersion has a higher chance of having its observed dispersion inflated. 
A dSph with a velocity dispersion between 4 and 10 $\kms$ is highly unlikely to be inflated by more than 30\% \citep{Minor10} (for an application of this method see \citet{Simon11, Martinez11}).
The objects we have found have observed velocity dispersions in this range. Assuming both objects are inflated by 30\%, their actual intrinsic velocity dispersion would be between $2.5-3.3~\kms$ and $7.1~\kms$ respectively, for the cold and velocity offset objects. 
These velocity dispersions are still much higher than $1~\kms$ (that is expected for a relaxed stellar system, i.e. a globular cluster).
It is unlikely that binary orbital motion alone can account for the large velocity dispersions inferred from this data set for both secondary populations. With multi-epoch data, we will be able test this hypothesis directly as was done for Segue 1 dSph \citep{Martinez11}. 

To assess the effect of tidal disruption from Ursa Minor we calculate the Jacobi Radius, $r_J$, and compare $r_J$ to the mean tidal radius estimated from our three-component analysis.
To calculate the Jacobi radius, we consider both an NFW \citep{Navarro97}  and a pseudo-isothermal (cored) profile for the halo of Ursa Minor. To set the  NFW density profile of Ursa Minor, we pick NFW scale radius $r_s = 1\ \rm kpc$ and estimate the density normalization $\rho_s$ using $M_{1/2}$ values from \citet{Wolf10} for a NFW profile. We find that if the actual distance of the center of the objects is equal to the projected distance from the center of Ursa Minor, then $r_J < r_t$.  If the objects are further than about $1\ \rm kpc$ away, then $r_J > r_t$ with the NFW profile. The situation for a pseudo-isothermal profile ($1/(r^2+r_0^2)$) with $r_0=300\ \rm pc$ is similar, with $r_J > r_t$ if the objects are further than about 1-2 kpc from the center of Ursa Minor. The $r_J$ estimates indicate that tides from Ursa Minor could have an effect on these objects even if they are protected by their own dark matter halos. The survival of globular cluster sized objects in dSphs has far-reaching implications 
for the density profile of the host halo \citep{Kleyna03, Goerdt06, Strigari2006, Cowsik09, Lora2012}. The objects we find are more extended and massive than the globular cluster sized objects considered in such work in the past. Thus these constraints will have to re-evaluated. 

Generically, the estimated high dispersions of these objects and their survival are facts at odds with each other. The age of Ursa Minor ($\sim 12\ \rm Gyr$) is much longer than the crossing time for stars inside Ursa Minor of $\sim 150\ \rm Myr$ (assuming a typical velocity of $10 \kms$). The crossing times for the stars in the cold and velocity offset object are $\sim 50\ \rm Myr$. These objects have had time to make multiple orbits around Ursa Minor and it is hard to see how they could have survived given the short crossing times unless  they have been recently captured by Ursa Minor and are now the process of tidal disrupted (which would account for the inflated velocity dispersion). However, this is not a likely scenario since Ursa Minor probably fell into the Milky Way early, between 8-11 Gyr \citep{Rocha11}, and capturing a large object after that is unlikely. It is more reasonable to assume that these objects have survived for long because they were protected by a dark matter halo of their own. The 
reality is probably 
more complicated: these objects may have their own dark matter halos and at the same time are being tidally disrupted. These implications are intimately tied to the dark matter halo of Ursa Minor and pinning down the properties of these objects would help to decipher if the dark matter halo of Ursa Minor has a cusp or a core. 

\section{Conclusion}

We have presented a method for finding multiple localized kinematically-distinct populations (stellar substructure) in line-of-sight velocity data.
In the the nearby dwarf spheroidal galaxy Ursa Minor, we have found two secondary populations: ``cold" and ``velocity offset (vo)" objects.
The estimated velocity dispersions are  $\sigma_{cold} = 4.25 \pm 0.75 \: \kms$ and $\sigma_{vo} = 9.25 \pm 1.25 \: \kms$, and the estimated mean velocities are $\ov_{cold} = -246.25 \pm 1.0 \: \kms$ and $\ov_{vo} = -258.0 \pm 1.5  \: \kms$.
They are located at $(0.25_{-0.06}^{+0.04}, -0.07_{-0.07}^{+0.03})\ \rm kpc$ (cold object) and $(-0.24 \pm 0.09, 0.23 \pm 0.02)\ \rm kpc$ (velocity offset object) with respect to the center of Ursa Minor. The location of the cold object matches that found earlier by \citet{Kleyna03}, but our results reveal that the velocity dispersion of this cold object could be large with a mean value close to $4\ \kms$. 
To assess the significance of our detections, we employed the Bayes Factor and information criteria $D_{KL}$ and DIC supplemented with the analysis of mock data sets with secondary populations, null hypothesis mock data sets and scrambled data sets.  The two secondary objects have $>98.5\%$ C.L. in all the model selection tests employed.

If the velocity dispersions are as large as our Bayesian analysis seems to indicate, then these objects are likely undergoing tidal disruption or are embedded in a dark matter halo. The two possibilities are not exclusive of each other. If these objects are dark matter dominated, this would be the first detection of a satellite of a satellite galaxy. 

As emphasized by \citet{Kleyna03} the presence of localized substructure has important implications for inner density profile of the dark matter halo of Ursa Minor. The shape of the inner profile (cusp or core) has important implications for the properties of the dark matter particle with cold dark matter model predicting a cuspy inner density profile. If the stellar substructure is hosted by its own dark matter halo, then it has further implications for dark matter models since this would likely be the smallest bound dark matter structure discovered. 

\section*{Acknowledgments}
This research was supported in part by the National Science Foundation Grant 0855462 at UC Irvine. This research was supported in part by the Perimeter Institute of Theoretical Physics during a visit by M.K. Research at Perimeter Institute is supported by the Government of Canada through Industry Canada and by the Province of Ontario through the Ministry of Economic Development and Innovation. 
G.D.M.~acknowledges support from the Wenner-Gren Foundations.
R.R.M.~acknowledges support from the GEMINI-CONICYT Fund, allocated to the project
N$^{\circ}32080010$, from CONICYT through project BASAL PFB-06 and the Fondo Nacional
de Investigaci\'on Cient\'ifica y Tecnol\'ogica (Fondecyt project N$^{\circ}1120013$).

\bibliographystyle{mnras}
\bibliography{bib}

\end{document}